\documentstyle[12pt,doublespace]{article}
\begin{document}

\vspace{0.20in}
\begin{center}
{\Large A Variational Estimate of the Binding Energy of Charge-Transfer
Excitons in the Cuprate Superconductors.}
\end{center}
\vspace{0.40in}
\begin{center}
C. Vermeulen and W. Barford.
\end{center}
\begin{center}
Department of Physics, The University of
Sheffield,  \\
\end{center}
\begin{center}
Sheffield, S3 7HR, United Kingdom.\\
\end{center}
\vspace{0.50in}
\begin{center}
Abstract
\end{center}
We present a variational estimate for the binding energy of a
Frenkel exciton in the insulating cuprate superconductors.
Starting from the three band Hubbard model we perform a canonical
transformation to O($t^2$), where $t$ is the bare nearest
neighbour copper-oxygen hopping integral. An effective Hamiltonian is then
derived to describe the hopping of the exciton through the copper oxide
plane. The critical parameter in the model is the nearest neighbour
copper-oxygen coulomb repulsion, $V$. It is found that a critical value of $V$
is needed to observe bound Frenkel excitons, and that these excitons have
the same symmetry as the parent copper orbital, $d_{x^2-y^2}$.
We determine the critical value of $V$ using a variational approach,
and attempt to fit the parameters of the model to known experimental results.
\pagestyle{empty}

\pagebreak

\noindent {\bf 1) Introduction.} \\
\\
Optical and Raman spectroscopy \cite{lui93}-\cite{perk93} indicate that
charge transfer excitons of the
inter-atomic kind ($d_{x^2-y^2}\rightarrow p_{x},p_{y}$) may be the lowest
lying excitations in the insulating
cuprate superconductors\cite{com1}.
There are also theoretical predictions that charge transfer fluctuations
play a crucial r$\hat{\mbox{o}}$le in determining the superconducting and
anomalous normal state
properties of the doped compounds \cite{bab92}-\cite{yun91}, although
experimental evidence for these excitations
in the doped phase is less clear.
Recent numerical work on the three band Hubbard Model with the nearest
neighbour
repulsion, $V$, has shown that charge transfer excitations are indeed the
lowest
lying excitations for a critical value of $V$\cite{verm94}.

We envisage the charge transfer exciton as a localised Frenkel exciton
consisting
of a copper hole excited onto its neighbouring oxygen sites, leaving behind
a vacant copper
site. The resultant bound particle-hole pair then delocalises through the
lattice, as illustrated in fig. 1. The potential energy of the exciton is
the nearest neighbour
Coulomb repulsion, $V$, as the oxygen hole has only
one neighbouring copper hole. This compares to the potential energy $2V$ for
an oxygen hole located amongst occupied copper sites. Thus the condition for
the exciton to lie in the charge transfer gap is that the potential
energy gained in forming an exciton must compensate the kinetic energy loss
 of binding the particle-hole pair.

 In this paper we present a variational estimate for the binding energy of a
 charge transfer exciton in the insulating state of the cuprate
superconductors.
 Starting from the three band model, which treats the copper and oxygen
orbitals
 on an equal footing, we perform a canonical transformation to O($t^2$), where
 $t$ is the copper-oxygen hybridization integral. We thus derive an effective
 low energy Hamiltonian which describes the charge dynamics of the copper-oxide
 planes. However, since we do not consider terms of O($t^4$) we neglect
 superexchange effects. The possible consequences of superexchange are
discussed
 in the conclusions. By keeping terms in the Hamiltonian which describe the
motion of
 the exciton we derive its energy in both a N$\acute{\mbox{e}}$el and a
 ferromagnetic background. This is therefore a variational estimate of the
 exciton energy. To calculate the exciton binding energy ({\it i.e.} the
energy from
 the bottom of the conduction band) we compare the exciton energy to the
minimum
 energy of the `single' particle-hole excitation energy. That is, the energy
 difference from the top of the valence band to the bottom of the
 conduction band.

 The plan of this paper is as follows. In the next section we perform the
canonical
 transformation on the three band model. In section 3 we use this to derive
 the excitonic Hamiltonian and calculate the energy of the exciton in the
insulating
 phase. Section 4 discusses the `single' particle Hamiltonians and
calculates the
 free particle-hole gap. In section 5 we calculate the condition for obtaining
 excitons in the charge transfer gap. Finally, in section 6 we attempt to
fit our
 model to the experimental results, and conclude.
\\
\\
\noindent {\bf 2) The Canonical Transformation.}
\\
\\
Our  starting point for the expansion is the unperturbed
   Hamiltonian
   \begin{equation}
   H = H_0 + H_t,
   \end{equation}
where
\begin{eqnarray}
H_0 &=& {\Delta \over 2} \sum_{ij\sigma}{(p^{\dagger}_{j\sigma}p_{j\sigma}
    -d^{\dagger}_{i\sigma}d_{i\sigma})}
        +  U_{d} \sum_{i} d^{\dagger}_{i\uparrow}d_{i\uparrow}
        d^{\dagger}_{i\downarrow}d_{i\downarrow}
        +  U_{p} \sum_{j} p^{\dagger}_{j\uparrow}p_{j\uparrow}
        p^{\dagger}_{j\downarrow}p_{j\downarrow}\nonumber\\
        &+&  V \sum_{<ij>\sigma\sigma^\prime}
\left(d^{\dagger}_{i\sigma}d_{i\sigma}
             p^{\dagger}_{j\sigma^\prime}p_{j\sigma^\prime}\right).
\end{eqnarray}
The perturbative part $H_{t}$ is defined as
\begin{equation}
 H_t = -t\sum_{<ij>\sigma} {(d^{\dagger}_{i\sigma}
p_{j\sigma} + h.c.), }
\end{equation}
\noindent where $i$ and $j$ are copper and oxygen sites
respectively,  $<ij>$ represents nearest neighbours and the operator
$d^{\dagger}_{i\sigma}(p^{\dagger}_{j\sigma})$ creates a copper (oxygen)
hole with
spin $\sigma$.  $\Delta$ is the charge-transfer energy, $U_d$
$(U_p)$ is the copper (oxygen) Coulomb repulsion,
and $V$ and $t$ are the
copper-oxide Coulomb repulsion and hybridisation, respectively.

The effective low energy Hamiltonian is formed using the unitary transformation
$\tilde{H}=e^SHe^{-S}$, where
$S^\dagger=-S$ \cite{mahan}. By expanding in $e^S$ it is trivial to show that
\begin{equation}
\tilde{H}=H_0+H_t+[S,H_0]+[S,H_t]+\frac{1}{2}[S,[S,H_0]]+O(t^3).
\end{equation}

To eliminate terms of order $t$ we define $S$ such that
$H_t+[S,H_0]=0$. The new Hamiltonian is then found by truncating the series
at terms greater than $O(t^2)$. Thus $\tilde{H}=H_0+H_{t^2}$, where
$H_{t^2}=\frac{1}{2}[S,H_t]$.
$\tilde{H}$ acts on the Hilbert space of $|\tilde{n}>$ where
$|\tilde{n}>=e^S|n>$
and $|n>$ is the basis of our original Hamiltonian.

By inspection we note that
\begin{equation}
[H_0,t\sum_{<ij>\sigma}(p_{j \sigma}^\dagger d_{i\sigma} - h.c.)]
=-t\sum_{<ij>\sigma}f_{ij\sigma}(d^\dagger_{i\sigma}p_{j\sigma}+ h.c.),
\end{equation}
where
\begin{equation}
f_{ij\sigma}=\Delta+(V-U_d)d^\dagger_{i\bar{\sigma}}d_{i\bar{\sigma}}
+(U_p-V)p^\dagger_{j\bar{\sigma}}p_{j\bar{\sigma}}+
V\sum_{<jj^{'}i>\sigma^{'}j\neq j^{'}}p^\dagger_{j^{'}\sigma^{'}}
p_{j^{'}\sigma^{'}}-V\sum_{<ii^{'}j>\sigma^{'}i\neq
i^{'}}d^\dagger_{i^{'}\sigma^{'}}
d_{i^{'}\sigma^{'}},
\end{equation}
and $j$ and $j^{'}$ neighbour $i$, and $i^{'}$ neighbours $j$.

It is relatively straight forward to show that
\begin{equation}
S=t\sum_{<ij>\sigma}(f_{ij\sigma})^{-1}(p^\dagger_{j\sigma}d_{i\sigma}- h.c.).
\end{equation}
\noindent Using the fact that $[f_{ij\sigma},d^\dagger_{i\sigma}
p_{j\sigma}]=[f_{ij\sigma},p^{\dagger}_{j\sigma}d_{i\sigma}]=0$ $H_{t^2}$ can
be
written as
\begin{eqnarray}
H_{t^2}=\frac{t^2}{2}\sum_{<ij>\sigma}\sum_{<kl>,\sigma^{'}}(f_{ij\sigma})^{-1}
&[&p^{\dagger}_{j\sigma} d_{i\sigma}d^{\dagger}_{k\sigma^{'}}p_{l\sigma^{'}}
+p^{\dagger}_{j\sigma}d_{i\sigma}p^{\dagger}_{l\sigma^{'}}d_{k\sigma^{'}}
\nonumber \\
&-&d^{\dagger}_{i\sigma}p_{j\sigma}d^{\dagger}_{k\sigma^{'}}p_{l\sigma^{'}}
-d^{\dagger}_{i\sigma}p_{j\sigma}p^{\dagger}_{l\sigma^{'}}d_{k\sigma^{'}}
+ h.c.].
\end{eqnarray}
\noindent The terms in equation (8) which change the number of oxygen or copper
holes are eliminated to yield
\begin{equation}
H_{t^2}=\frac{t^2}{2}\sum_{<ij>\sigma}\sum_{<kl>,\sigma^{'}}(f_{ij\sigma})^{-1}
[p^{\dagger}_{j\sigma} d_{i\sigma}d^{\dagger}_{k\sigma^{'}}p_{l\sigma^{'}}
-d^{\dagger}_{i\sigma}p_{j\sigma}p^{\dagger}_{l\sigma^{'}}d_{k\sigma^{'}}
+ h.c].
\end{equation}
Although our Hamiltonian has been simplified it
is still very complicated,
containing residual many body effects resulting from copper-oxygen coulomb
repulsion
and the effects of onsite copper and oxygen coulomb repulsion.
\\
\\
\noindent {\bf 3) The Excitonic Hamiltonian.}
\\
\\
In constructing the excitonic Hamiltonian we first consider the part of
equation
(9) which hops the oxygen hole around a single copper-oxygen plaquette.
This may be written in the form
\begin{equation}
H=\sum_ih_i,
\end{equation}
where
\begin{equation}
h_i=\frac{t^2}{(\Delta+V)}\sum_{<jj^{'}>\sigma}p^{\dagger}_{j^{'}\sigma}
p_{j\sigma}.
\end{equation}
Thus when the hole hops to a neighbouring oxygen site via an empty copper site
it costs an
energy $\frac{t^2}{(\Delta+V)}$. The symmetry of this Hamiltonian is exploited
by noting that the only non-zero solution of $h_i$
is the constant phase solution $P_{i\sigma}
=\frac{1}{2}(p^{\dagger}_{1\sigma}+p^{\dagger}_{2\sigma}+p^{\dagger}_{3\sigma}+
p^{\dagger}_{4\sigma})$, where the indices on the oxygen operators stand for
the
four neighbours of the copper site at $i$. From this it can be inferred that
our
exciton has the same symmetry as the parent copper site, $d_{x^2-y^2}$, which
is the $A_{1g}$ symmetry in
Raman notation. $h_i$ can be rearranged into the more compact form,
$h_i=\frac{t^2}
{(\Delta+V)}P^{\dagger}_{i\sigma}P_{i\sigma}$.
The full Hamiltonian is then derived by only keeping terms in equation (9)
which
hop this entity through the `sea' of Cu$^{2+}$ copper sites.
We then arrive at the final Hamiltonian
\begin{eqnarray}
H_e&=&\sum_{<ik>\sigma\sigma^{'}}\alpha_{\sigma\sigma^{'}}
E^{\dagger}_{i\sigma}E_{k\sigma^{'}}+\sum_{<ik>\sigma}\beta
e^{\dagger}_{i\sigma}e_{k\sigma}
+\sum_{i\sigma}\gamma E^{\dagger}_{i\sigma}
E_{i\sigma}      \nonumber \\
&+&\sum_{<ik>\sigma}\delta E^{\dagger}_{i\sigma}E_{i\sigma}d^{\dagger}
_{k\bar{\sigma}}d_{k\bar{\sigma}}
+\sum_{<ik>\sigma}\epsilon
e^{\dagger}_{i\bar{\sigma}}E_{i\sigma}d^{\dagger}_{k\sigma}d_{k\bar{\sigma}}
+ (V+\Delta),
\end{eqnarray}
where the zero point energy is given as the spin degenerate ground state of
one hole per copper site.

The operator $E^{\dagger}_{i\sigma}=P^{\dagger}_{i\sigma}d_{i\sigma}$
creates a spin $0$
exciton on site $i$, while $e^{\dagger}_{i\sigma}
=P^{\dagger}_{i\sigma}d_{i\bar{\sigma}}$ spin flips and creates a spin $1$
exciton on site $i$. The first two terms in $H_e$ hop these excitons through
the lattice. The remaining terms in $H_e$ are self energy terms, which include
spin flips.
 $i$ and $k$ are nearest neighbour {\it copper} sites. The coefficients
are
\begin{eqnarray}
 \alpha_{\sigma\sigma}&=&\frac{t^{2}(\Delta-3V)}{4\Delta(\Delta+V)} ,
 \hspace{1cm}
\alpha_{\sigma\bar{\sigma}}=\frac{t^{2}V(3V-5\Delta)+U_p(\Delta-3V)}
 {4\Delta(\Delta+V)(\Delta+U_p-V)}, \nonumber \\
 \beta&=&\frac{t^{2}}{4(\Delta+U_p-V)} ,\hspace{1cm}
 \gamma=\frac{t^{2}(9\Delta-6V)}{\Delta(\Delta+V)}  \nonumber \\
 \mbox{\noindent
and}\hspace{1cm}\delta&=&-\epsilon=\frac{t^{2}(U_p+U_d-2V)}{4(\Delta+V-U_d)(
\Delta+U_p-V)}.
 \end{eqnarray}
The fact that $\delta=-\epsilon$ is due to the effects of exchange.

With this Hamiltonian we now wish to construct a variational estimate for
the energy of the exciton.
This is done by considering the dispersion of the exciton through either a
ferromagnetic or a N$\acute{\mbox{e}}$el background.
In the ferromagnetic case we set
$\alpha_{\sigma\sigma^{'}}=\alpha_{\sigma\sigma}$ and
$\beta=\delta=\epsilon=0$.
The variational wave function is then
\begin{equation}
|\phi_F>=\frac{1}{\sqrt{N_{cu}}}\sum_{l}e^{i\underline{k}.\underline{l}}
E^{\dagger}_{l\uparrow}\prod_{j}d^{\dagger}_{j\uparrow}|0>,
\end{equation}
with an energy
\begin{equation}
E^F_{e}(\underline{k})=(V+\Delta)+\frac{t^2(\Delta-3V)}{2\Delta(\Delta+V)}[c
osk_x+cosk_y]+
\frac{t^{2}(9\Delta-6V)}{\Delta(\Delta+V)}.
\end{equation}
Similarly, the variational energy of an exciton in the
N$\acute{\mbox{e}}$el background is found by setting
$\alpha_{\sigma\sigma^{'}}=\alpha_{\sigma\bar{\sigma}}$ and $\beta=\epsilon=0.$
 Thus,
\begin{equation}
|\phi_N>=\frac{1}{\sqrt{N_{cu}}}\sum_{l\sigma}e^{i\underline{k}.\underline{l}}
E^{\dagger}_{l\sigma}|\mbox{N$\acute{\mbox{e}}$el}>,
\end{equation}
with
\begin{eqnarray}
E^N_{e}(\underline{k})&=&(V+\Delta)+\frac{t^{2}V(3V-5\Delta)+U_p(\Delta-3V)}
{2\Delta(\Delta+V)(\Delta+U_p-V)}[cos(k_x)+cos(k_y)]  \\
&+&\frac{t^{2}(U_p+U_d-2V)}{(\Delta+V-U_d)(\Delta+U_p-V)}
+\frac{t^{2}(9\Delta-6V)}{\Delta(\Delta+V)}.
\end{eqnarray}

In the limit of $U_p=\infty$
$E_{e}^{N}(\underline{k})=E_{e}^{F}(\underline{k})$,
and for $U_p<\infty$
$E_{e}^{N}(min)<E_{e}^{F}(min)$ in our allowed parameter range of
$0<V<(\Delta+U_p)$. It is therefore our intention to consider the scenario
of an exciton moving in the Ne$\acute{\mbox{e}}$l background, which is also the
experimentally relevant case.
The minimum energy of the exciton can be expressed in terms of the potential
energy and an effective kinetic energy. For the case of $U_d=\infty$ this is
written as
\begin{eqnarray}
E_{exc}^N&=&(V+\Delta)-\frac{t^2}{\Delta}f_{exc}^N(V,U_p,\Delta),
\end{eqnarray}
where
\begin{eqnarray}
f_{exc}^N(V,U_p)&=&-\left(\frac{V(9V-21\Delta)+U_p(10\Delta-9V)+8\Delta^2}
{(\Delta+V)(\Delta+U_p-V)}\right).
\end{eqnarray}
\\
\\
\noindent {\bf 4) Single particle-hole Hamiltonian.}
\\
\\
As we wish to determine the condition for a bound exciton in the charge
transfer
gap, we need to consider the motion of an unbound particle-hole pair in order
to find the single particle gap energy.
In determining the energy of the gap for a free particle hole pair excitation
 we intend to study separately the energies of an added hole and
 a removed hole from the insulating phase. The gap energy is then given as
 $E_{gap}=E(N_{cu}+1)+E(N_{cu}-1)-2E(N_{cu})$, where $N_{cu}$ is the number of
 copper sites. We start by considering the energy of a removed hole, and then
 consider the more complicated scenario of an added hole.
\\
\\
4.1) Removed Hole.
\\
\\
The Hamiltonian which describes the motion of a removed hole amongst $N_{cu}-1$
occupied copper sites is\cite{long89}
\begin{equation}
H_{v}=\frac{\Delta}{2}-\frac{4t^2}{\Delta}\sum_{i\sigma}c^{\dagger}_{i\sigma
}c_{i\sigma}
+\frac{t^2}{\Delta}\sum_{<ik>\sigma}c^{\dagger}_{k\sigma}c_{i\sigma}+\sum_{i
}\frac{4t^2}{(\Delta+V)},
\end{equation}
where $c^{\dagger}_{i\sigma}$ creates a particle (destroys a hole {\it i.e.}
$c_{i\sigma}^{\dagger}=d_{i\sigma}$) on site $i$,
and the zero point energy is as before.
The Nagaoka theorem\cite{nag66} informs us that in the strong coupling limit
the maximum bandwidth for the dispersion of the empty site is obtained in a
ferromagnetic background.
The $S=(N_{cu}-1)$ branch of the spectrum is then
\begin{equation}
E^{v}(\underline{k})=\frac{\Delta}{2}+\frac{2t^2}{\Delta}[cos(k_x)+cos(k_y)]
-\frac{4t^2}{\Delta}+\frac{8t^2}{(\Delta+V)}.
\end{equation}
The {\it minimum} energy of the empty site, at $\underline{k}=(\pi,\pi)$,
is the {\it top} of the valence band
\\
\\
4.2) Added Hole.
\\
\\
    The Hamiltonian for the added hole follows the analysis of Barford
\cite{bar89}.
In that paper the same canonical transformation, equation (4), was
performed on the
Hamiltonian and the effective Hamiltonian was then diagonalised for the
$i^{\mbox{th}}$ plaquette of one hole on the oxygen sites. This was done
for the case $U_p=0$.
For a general $U_p$ the Hamiltonian is given as
\begin{eqnarray}
H_c&=&(2V+\frac{\Delta}{2})-(t_1-t_2)\sum_{i}S_{i0}^{\dagger}.S_{i0}
\nonumber \\
&+&t_3\sum_{i,k}S_{ik}^{\dagger}.S_{ik}
+t_1\sum_{i\delta}T_{i\delta}^{\dagger}.T_{i\delta}
-\sum_{i}\left(\frac{4t^2}{\Delta}-\frac{4t^2}{(\Delta+V)}\right).
\end{eqnarray}
The zero point energy is again that of a half filled plane
$\left(E(N_{cu})=\frac{-N\Delta}{2}-\frac{4t^2N}{(\Delta+V)}\right)$.
 $t_1=\frac{4t^2}{\Delta}$,
 $t_2=\frac{8t^2}{(\Delta+2V-U_d)}$ and
$t_3=\frac{2t^2U_p}{\Delta(\Delta+U_p)}$.
$i$ represents all the plaquettes with an added hole. The operators are defined
as
\begin{eqnarray}
S^{\dagger}_{ik}&=&\frac{1}{\sqrt{2}}(P^{\dagger}_{ik\uparrow}
d^{\dagger}_{i\downarrow}-P_{ik\downarrow}^{\dagger}d^{\dagger}_{i\uparrow})
\hspace{0.5cm}k=0,\underline{+}\frac{\pi}{2},\hspace{0.3cm}\pi \\
P_{i,k\sigma}^{\dagger}&=&\frac{1}{2}\sum_{l}e^{ikl}p^{\dagger}_{l\sigma} \\
T_{i0}&=&\frac{1}{\sqrt{2}}(P^{\dagger}_{i\uparrow}
d^{\dagger}_{i\downarrow}+P_{i\downarrow}^{\dagger}d^{\dagger}_{i\uparrow}), \\
T_{i+}&=&P^{\dagger}_{i\uparrow}d^{\dagger}_{i\uparrow} \\
\mbox{\noindent
and}\hspace{2cm}T_{i-}&=&P^{\dagger}_{i\downarrow}d^{\dagger}_{i\downarrow}.
\end{eqnarray}
$S_{i0}$ is the `Zhang-Rice' singlet operator.
The factor $\frac{8t^2}{(\Delta+2V-U_d)}$ arises from virtual $\mbox{Cu}^+$
hopping,
while the remaining terms are accounted for by virtual $\mbox{Cu}^{3+}$
hopping.
The additional complication to the $U_p=0$ case is the existence of
non-zero anti symmetric
eigenvalues for the singlet Cu$^{2+}$O$^{-}$ pair. The oxygen hole can bond
in a singlet configuration with its neighbouring copper hole and pick up a
phase change as it hops around the plaquette.
The motion of the added hole is complicated because a triplet state on
the $i^{\mbox{th}}$ site is, in general, not orthogonal to a singlet or triplet
state on the neighbours of $i$. It is also unclear how the anti symmetric
singlet
states hybridise with neighbouring sites. It is clear, however, that their
bands are
relatively flat and lie between the singlet and triplet symmetric states.
Consequently we do not intend to consider these states in our equation.

We now wish to construct a variational estimate for the minimum energy of this
Hamiltonian.
To do this we again invoke the Nagaoka theorem\cite{nag66} which tells us
that the
maximum bandwidth for the doped hole occurs in a ferromagnetic background.
Given this the most obvious choice of variational wave function is that of
a singlet moving in a ferromagnetic background. This has the Bloch wave
function
\begin{eqnarray}
|\alpha_{k}>&=&\frac{1}{\sqrt{N_{cu}}}\sum_{l}e^{i\underline{k}
.\underline{l}}S^{\dagger}_{l0}
d_{l\uparrow}\prod_{j}d^{\dagger}_{j\uparrow}|0>.
\end{eqnarray}
This can scatter into a triplet $S_z=0$ state of the form
\begin{eqnarray}
|\beta_{k}>&=&\frac{1}{\sqrt{N_{cu}}}\sum_{l}e^{i\underline{k}.\underline{l}
}T^{\dagger}
_{l0}d_{l\uparrow}\prod_{j}d^{\dagger}_{j\uparrow}|0>,
\end{eqnarray}
and into a triplet $S_z=1$ state of the form
\begin{eqnarray}
|\gamma_{k}>&=&\frac{1}{2\sqrt{N_{cu}}}\sum_{<ll^{'}>}e^{i\underline{k}.
\underline{l}}
T^{\dagger}_{l^{'}+}d_{l^{'}\uparrow}d^{\dagger}_{l\downarrow}
d_{l\uparrow}\prod_{j}d^{\dagger}_{j\uparrow}|0>.
\end{eqnarray}
These three wave functions give the following overlaps:
\begin{eqnarray}
<\alpha_{k^{'}}|\beta_{k}>&=&-\delta_{ kk^{'}}\frac{1}{4}[cos(k_x)+cos(k_y)],
\\
<\alpha_{k^{'}}|\alpha_{k}>&=&\delta_{kk^{'}}(1+\frac{1}{4}[cos(k_x)+cos(k_y
)]), \\
<\beta_{k^{'}}|\beta_{k}>&=&\delta_{kk^{'}}(1+\frac{1}{4}[cos(k_x)+cos(k_y)])
\\
\mbox{\noindent
and}\hspace{2cm}<\beta_{k^{'}}|\gamma_{k}>&=&-<\alpha_{k^{'}}|\gamma_{k0}>=\
frac{1}{\sqrt{2}}
\delta_{kk^{'}}.
\end{eqnarray}
The variational energy is given by $E_c=\frac{<\alpha_{k}
|H_c|\alpha_{k}>}{<\alpha_{k}|\alpha_{k}>}$.

Now, the Hamiltonian acting on the state $|\alpha_{k}>$ gives
\begin{eqnarray}
H_c|\alpha_k>=&-&(t_1-t_2-t_3)|\alpha_k>-(t_1-t_2-t_3)\frac{1}{4}(cosk_x+cos
k_y)|\alpha_k>    \nonumber \\
&+&\frac{t_1}{4}(cosk_x+cosk_y)|\beta_k>-\frac{t_1}{\sqrt{2}}|\gamma_k>,
\end{eqnarray}
so usng the overlaps, equations (32) to (35), $E_c$ is given by
\begin{equation}
E_c=(2V+\frac{\Delta}{2})+\frac{(-(t_1-t_2-t_3)-(t_1-t_2-t_3)\frac{e_k}{4})(
1+\frac{e_k}{4})
  +\frac{t_1e^2_k}{4^2}+\frac{t_1}{2}}{(1+\frac{e_k}{4})}-\frac{8t^2}{\Delta}
  +\frac{8t^2}{(\Delta+V)},
\end{equation}
where $e_k=[cos(k_x)+cos(k_y)]$. This value of the energy agrees very well with
numerical calculations on $4x4$ clusters\cite{long89}.

In what now follows we will assume that only $\mbox{Cu}^+$ virtual excitations
are present. This is achieved by setting $U_d$ to infinity, i.e. $t_2=0$. $U_d$
is
known to be large in the cuprates so this assumption has physical
validity.
The energy of the excitation gap is then simply the distance between the
top of the valence band and the bottom of the conduction band.
The top of the valence band is at $\underline{k}=(\pi,\pi)$ and the bottom
of the
conduction band is given at $\underline{k}=(0,0)$.
Thus the minimum gap energy is given as $E_{gap}=E^c(0,0)+E^v(\pi,\pi)$
($E(N_{cu})=0$) Although this is not a momentum conserving transistion, we take
this as the gap energy as we require an upper bound on the value of $V$ for
obtaining a bound exciton.
The gap energy can be rewritten in the form
\begin{equation}
  E_{gap}=(2V+\Delta)-\frac{t^2}{\Delta}f_{gap}(V,U_p,\Delta)
\end{equation}
where
\begin{equation}
f_{gap}(V,U_p)=4+\frac{16V}{(\Delta+V)}-\frac{3U_p}{(\Delta+U_p)}.
\end{equation}
\\
\\
\noindent {\bf 5) Exciton Binding Energy.}
\\
\\
Having calculated the minimum energy of the exciton in the anti
ferromagnetic phase,
and the minimum distance between the top of the valence band and the bottom of
the conduction band we can now calculate the binding energy of the exciton.
This
is defined as
\begin{eqnarray}
E_{bin}&=&E_{gap}-E_{exc} \nonumber \\
&=&V+\frac{t^2}{\Delta}(f_{gap}-f_{exc}).
\end{eqnarray}
The condition $E_{bin}=0$ determines the critical value of $V$ for there to
be an
exciton in the charge transfer gap.
This is illustrated
in fig.2 where the critical value of
$\tilde{V}$ is plotted against $\tilde{\Delta}$ (where
$\tilde{V}=\frac{V}{\tilde{t}}$, $\tilde{\Delta}=\frac{\Delta}{\tilde{t}}$
and $\tilde{t}=\frac{t^2}{\Delta}$)
for various values of $U_p$.
We emphasise that this is an {\it upper} bound on the critical value of
$V$, as we
have underestimated the gap expected from an optical transistion,
and the energy of the exciton is variational.
Equation (40) illustrates the competing
effects of the potential energy gain (-$V$) and the kinetic energy loss
$\frac{t^2}{\Delta}(f_{gap}-f_{exc})$ in forming a bound exciton. Notice that
the kinetic energy scale is $\tilde{t}=\frac{t^2}{\Delta}$.
In fig.3 we plot the binding energy $E_{bin}$ against $\tilde{V}$ for
$\tilde{\Delta}=9\tilde{t}$ which can be interpreted as a $\Delta$ of $3eV$ and
$t=1.0eV$.

The effective mass is defined as
$\frac{1}{m^*}=\frac{1}{\hbar^2}\frac{\partial^2E}{\partial k^2}$.
Thus, the ratio of the effective masses of the exciton is given as
\begin{equation}
\frac{m_{ex}}{m_{hole}}=\frac{\partial^2E_{hole}}{\partial
k^2}/\frac{\partial^2E_{ex}}
{\partial k^2},
\end{equation}
{\it i.e.} their mass ratio is essentially just the ratio of their band widths.
The  free hole
band width is given as $E_c(\pi,\pi)-E_c(0,0)$.
The exciton band width is $E_{exc}(\pi,\pi)-E_{exc}(0,0)$.
This gives a mass ratio of
\begin{equation}
\frac{m_{ex}}{m_{hole}}=-\frac{(4\Delta+2U_p)(\Delta+U_p-V)(\Delta+V)}
{2(\Delta+U_p)[(V(3V-5\Delta)+U_p(\Delta-3V)]}.
\end{equation}
\\
\\
\noindent {\bf 6) Discussions and Conclusions.}
\\
\\
The theory presented in this paper predicts the binding energy of a charge
transfer exciton with
$d_{x^2-y^2}$ symmetry (A$_{1g}$ in Raman notation) as a function of $V$,
$\Delta$ and $U_p$. We now compare these predictions to the experimental data.
Recent optical absorption experiments \cite{perk93} give a value of the gap
energy as $E_{gap}\sim1.7eV$,
and a recent Raman scattering study of insulating cuprates \cite{lui93}
gives the binding energy
of the $A_{1g}$ exciton as $E_{bin}\sim0.2eV$.
The value of the effective hopping amplitude $\tilde{t}$ may be estimated from
measurements of the superexchange interaction \cite{aep89,end88}, $J$, since
$J=\frac{4\tilde{t}}{U_d}\sim0.13eV$ and $U_d\sim10eV$ from photoemmission
data. This gives us an estimate for $\tilde{t}\sim0.6eV$.
For {\it fixed} $U_p$
we therefore have two equations (19) and (38) with two unknowns and so $V$
and $\Delta$ can
be evaluated.
This proceeds as follows. Firstly $\tilde{t}$ is eliminated from equations
(19) and (38)
to give the relation
\begin{equation}
(E_{exc}^N-(V+\Delta))f_{gap}(V,U_p)=(E_{gap}-(2V+\Delta))f_{exc}^N(V,U_p).
\end{equation}
This gives an expression for $V$ and $\Delta$ which can then be reinserted into
the expressions for the energy of the gap to give the value of $\tilde{t}.$

For the case of $U_p=0$ equation (43) is a quintic polynomial in $V$ and
$\Delta$.
 This is solved using the bisection method to give $V\sim2.7eV$ and
$\Delta\sim3.5eV$.
This gives the bare hopping matrix element as $t\sim1.4eV$.
Although these parameters are quite reasonable the
disappointing feature is that the value of $V$ is large, and greater than
$U_p$.
This is probably unphysical, although direct oxygen-oxygen hopping would
be expected to screen $U_p$.
It should also be remembered that this calculation predicts an upper bound
on the value of $V$ required
for excitons.
The mass of the exciton is calculated
to be $m_{exc}=0.39m_{hole}$.

For the case $U_p=\infty$ equation (43) is a quadratic in $\Delta$ which can be
solved using the usual quadratic formula. This gives a value of $V\sim5.7eV$,
$\Delta\sim0.4eV$ and $t=0.5eV$. This time {\it all} the parameters are
inconsistent with the values
currently accepted by most to be those of the cuprate superconductors,
i.e. $V\sim1eV$, $\Delta\sim3eV$, $t\sim1.5eV$ \cite{tun91}. The exciton mass
is $m_{exc}=0.36m_{hole}$.

In conclusion, we have performed a canonical transformation of the three band
Hubbard model up to O($t^2$) in the bare hopping amplitude and derived a
variational estimate for the energy of an exciton and that of a free
particle-hole pair moving through the copper oxide plane. We find that the
kinetic energies of our particles are rescaled by an effective hopping
amplitude
$\tilde{t}=\frac{t^2}{\Delta}$, and that a critical value of the nearest
neighbour Coulomb repulsion is required to observe bound excitons. This
critical
value of $V$ is also strongly dependent on the value of $U_p$, the
oxygen-oxygen
Coulomb repulsion.
The exciton consists of an oxygen hole in a constant phase sum of the orbitals
tied to an empty copper site. It therefore has $d_{x^2-y^2}$
 symmetry . Such an
excitation is electric dipole forbidden, but an electric-field induced dipole
transition is allowed if the polarized light is parallel to the electric field
\cite{falck92}. Unfortunately such an experiment has not yet been performed.

In the Raman spectroscopy terminology the exciton has A$_{1g}$ symmetry and is
Raman active. These excitations have been observed, and by fitting the
experimental data to our theory we have found an upper bound for $V$ of $2.7eV$
assuming $U_p=0$, or $5.7eV$ assuming $U_P=\infty$. The latter result is
undoubtedly unrealistic.

In this paper we have not considered the effects of direct oxygen-oxygen
hopping, $t_{pp}$. Including this would have two effects. Firstly it would
screen $U_p$. Secondly it would render a non-bonding orbital bonding with
$d_{3z^2-r^2}$ symmetry.
The energy of this exciton is lowered while the $d_{x^2-y^2}$
symmetry state is raised.
An exciton with $d_{3z^2-r^2}$ symmetry is
observed in both optical and Raman spectroscopy with a binding energy of
about $1.7eV$.
In addition to neglecting $t_{pp}$ we have also neglected the superexchange
interaction.
This interaction would be expected to narrow the valence and conduction
bands and hence widen
the single particle energy gap. However, it is unlikely to effect our
estimate of the
exciton energy, as we assume that it propagates through a N$\acute{mbox{e}}$el
background. Consequently we expect that the superexchange term would {\it
reduce}
the value of $V$ required for bound excitons from the value calculated by
this theory.
\begin{center}
{\bf Acknowledgements.}\\
\end{center}
W. B. would like to thank E. R. Gagliano and J. Lorenzana for useful
discussions.
C.J.V. would like to acknowledge the support of the
University of Sheffield for providing him with a scholarship.

\pagebreak

\begin{figure}
{\bf Figure 1:} An exciton in a N$\acute{mbox{e}}$el background, mving in the
copper oxide plane of a cuprate superconductor.
\\
\\
{\bf Figure 2:} These curves show the boundary between a bound exciton moving
in the N$\acute{\mbox{e}}$el background and a free particle-hole pair for
several values of $U_p$.
Above the line a bound exciton has the lower energy.
\\
\\
{\bf Figure 3:} The binding energy of the exciton and gap energy for (a)
$U_p=0$ and
(b) $U_p=\infty$, in units of $\tilde{t}$, as a function of $\tilde{V}$.
\end{figure}


\begin{thebibliography}{10}
\bibitem{lui93} Ran Liu et al, Phys.\ Rev.\ Lett.\,{\bf71}, 3709 (1993).
\bibitem{falck92} J. P. Falck, M. A. Kastner and R. J. Birgeneau, \ Phys. \
Rev. \
Lett. \ {\bf 69} 1109 (1992).
\bibitem{perk93} J. D. Perkins et al, Phys.\ Rev.\ Lett.\ {\bf71} 1621 (1993).
\bibitem{com1}
An alternative explanation is that these low lying excitations may be
intra-atomic
excitations ($d_{x^2-y^2}\rightarrow d_{xy},d_{3z^2-r^2},d_{xz},d_{yz}$)
\bibitem{bab92} V. S. Babichenko and M. N. Kiselev, J.\ Moscow\ Phys.\ soc.\ 2,
311 (1992).
\bibitem{yu91} J. Lorenzana and L. Yu, Phys.\ Rev.\ B \ {\bf 43} 11474 (1991).
\bibitem{Nun88} M. D. Nu$\tilde{\mbox{n}}$ez Regueiro and A. A. Aligia,
Phys.\ Rev.\ Lett.\ {\bf 61} 1889 (1988).
\bibitem{suj} T.Tsuji, O. Narikiyo and K. Miyake, Dept. of Material Physics,
Faculty of Engineering Science, Osaka Univ., Osaka 560, Japan.
\bibitem{gab84} M. Gabowski and L. J. Sham, Phys.\ Rev.\ B\ {\bf29} 6132
(1984).
\bibitem{yun91} Yunkyu Bang, G. Kotliar, C. Castellani, M. Grilli and R.
Raimondi, Phys.\ Rev.\ B\ {\bf43} 13724 (1991).
\bibitem{verm94} C. Vermeulen, W. Barford, E. R. Gagliano, Univ. of Sheffield,
Sheffield, U.K. (preprint).
\bibitem{mahan} G. D. Mahan, Many Particle Physics 2$^{\mbox{nd}}$ Ed. Plenum
Press.
\bibitem{long89} M. W. Long and W. Barford, Rutherford Appleton Laboratory
Report
RAL 89-053 (1989).
\bibitem{nag66} Y. Nagaoka, Phys.\ Rev.\ {\bf147} 392 (1966).
\bibitem{bar89} W. Barford, J. Phys. (Cond. Matt.) {\bf 2} 2965 (1990).
\bibitem{aep89} G. Aeppli et al, Phys.\ Rev.\ Lett.\ {\bf 62} 2052 (1989).
\bibitem{end88} Y. Endoh et al, Phys. \ Rev.\ B {\bf 37} 7443 (1988).
\bibitem{tun91} G. A. Sawatzky. 1991,
High Temperature Superconductivity. ed. D. P. Tunstall and W. Barford,
Publishers: Adam Hilger. (Bristol)
\end{thebibliography}
\end{document}